\documentclass[aps,prb,twocolumn,groupedaddress,showpacs,amstex]{revtex4}
\usepackage{graphicx}

\newcommand{\bone}{\textbf{B1}}
\newcommand{\btwo}{\textbf{B2}}
\newcommand{\bthree}{\textbf{B3}}

\begin{document}

\title{Quantum phases of correlated electrons in
artificial molecules under magnetic fields}

\author{Devis Bellucci}
\email[]{dbellucci@unimore.it}
\homepage[]{www.nanoscience.unimore.it}
\affiliation{CNR-INFM  National Research Center on
nanoStructures and bioSystems at Surfaces (S3) and \\
Dipartimento di Fisica, Universit\`a degli Studi di Modena e
Reggio Emilia, Via Campi 213/A, 41100 Modena, Italy}
\author{Massimo Rontani}
\affiliation{CNR-INFM National Research Center on nanoStructures
and bioSystems at Surfaces (S3) and \\
Dipartimento di Fisica, Universit\`a degli Studi di Modena e Reggio
Emilia, Via Campi 213/A, 41100 Modena, Italy}
\author{Guido Goldoni}
\affiliation{CNR-INFM National Research Center on
nanoStructures and bioSystems at Surfaces (S3) and \\
Dipartimento di Fisica, Universit\`a degli Studi di Modena e
Reggio Emilia, Via Campi 213/A, 41100 Modena, Italy}
\author{Elisa Molinari}
\affiliation{CNR-INFM National Research Center on
nanoStructures and bioSystems at Surfaces (S3) and \\
Dipartimento di Fisica, Universit\`a degli Studi di Modena e
Reggio Emilia, Via Campi 213/A, 41100 Modena, Italy}

\date{\today}

\begin{abstract}
We  investigate the stability of few-electron quantum phases in
vertically coupled quantum dots  under a magnetic field of arbitrary
strength and direction. The orbital and spin stability diagrams of
realistic devices containing up to five electrons, from strong to
weak inter-dot coupling, is determined. Correlation effects and
realistic sample geometries are fully taken into account within the
Full Configuration Interaction method. In general, the magnetic field drives the
system into a strongly correlated regime by modulating the
single-particle gaps. In coupled quantum dots different components
of the field, either parallel or perpendicular to the tunneling
direction, affect single-dot orbitals and tunneling energy,
respectively. Therefore, the stability of the quantum phases is
related to different correlation mechanisms, depending on the field
direction. Comparison of exact diagonalization results with simple models allows to identify the
specific role of correlations.
\end{abstract}


\maketitle

\section{INTRODUCTION}

Due to three-dimensional confinement on length scales comparable to
the De Broglie wavelength, the electronic properties of
semiconductor quantum dots (QDs) show several similarities with
those of atoms, the most significant ones being their discrete
energy spectrum and the resulting shell
structure.\cite{ashoori,tarucha1} Fine structure due to the exchange
interaction (Hund's rule)\cite{tarucha1,RontaniAPL,tarucha2} and
Kondo physics \cite{gordon,Cronenwett98} have also been predicted
and demonstrated. QDs are therefore regarded as
\textit{artificial
atoms}.\cite{Maksym,kastner,jacak,reimann1,chakraborty} These
systems have stimulated the investigation of the fundamentals of few
body physics in semiconductors, since the number of carriers,
electrons or holes, in QDs can be controlled very accurately, and
almost all relevant parameters influencing their strongly correlated
states, such as confinement potential or the coupling with static
magnetic fields, can be tailored in the experiments, driving the
system between very different regimes.

The analogy between artificial and natural atoms is extended to {\em
artificial molecules} (AMs) by realizing tunnel-coupled
devices.\cite{rontani2,blick,schedelbeck,Austing98,holleitner,Ota05}
The inter-dot tunnelling introduces a new energy scale which may be
comparable to other energy scales, namely the single-particle (SP)
confinement energy, the carrier-carrier interaction, and the
magnetic confinement energy. In AMs it is possible to tune the
coupling among QDs basically at will, thus exploring different
molecular bonding regimes;\cite{rontani} this is an intriguing
option with respect to natural molecules, since for the latter the
inter-nuclear coupling is almost fixed by the balance between
nuclear repulsion and the electrostatic attraction mediated by
valence electrons.

There is a number of different techniques to fabricate AMs.
Laterally coupled AMs\cite{waugh,Oosterkamp,Fujisawa,Ladriere,Petta}
are obtained by creating an electrostatic confinement in a
semiconductor heterostructure, such as a doped heterojunction, by
means of the photo-lithographic patterning of metallic gates,
deposited on the surface of the heterostructure. Another way to
realize AMs is by the Stransky-Krastanov mechanism leading to the
formation of self-assembled QDs with nm-scale confinement with
similar sizes and regular shapes. In a stack of several layers, the
formation of QDs in the top layer, coupled through tunneling with
the ones in the underlying layer, has been recently
demonstrated.\cite{bayer,fafard}

In the following we shall investigate electronic properties of
vertically coupled AMs realized starting from a triple barrier
heterostructure to form a mesa pillar, possibly using a combination
of dry and wet
etching.\cite{Austing98,Schmidt97,amaha,pi,Ono,ancilotto,austing,rontani}
Typically, the effective diameter for electron conduction is smaller
than the geometrical diameter of the top contact. For example, with
a top contact $\approx$ 0.5 $\mu$m diameter, it is estimated that
the dot diameter is $\approx$ 0.1 $\mu$m when the dot contains a few
electrons. Still, in this class of devices the depletion potential
in the lateral direction realizes a much softer potential than in
the growth (vertical) direction, where electrostatic confinement is
induced by the band mismatch of the layers in the heterostructure.
In this vertical geometry transport occurs perpendicularly to the
plane of the dots, in response to an applied voltage between source
and drain contacts. The strength of the coupling can be modulated by
the width of the barrier separating the QDs. In addition, a circular
Schottky gate is typically placed around the body of the pillar to
control the charging of the QDs.

In AMs, as well as in single QDs, electronic states can be
manipulated by a \emph{vertical} magnetic field $B_\perp$,
perpendicular to the plane of the QDs (parallel to the growth
direction), which drives the system from a low-correlation
(low-field) regime to a strongly correlated (high-field) regime by
varying the SP splittings.\cite{reimann1} Recently, non trivial
transitions between different quantum states induced by $B_\perp$
and/or by varying the inter-dot barrier width $L_{b}$ have been
predicted and demonstrated in AMs with different coupling
regimes.\cite{rontani99,bart00,rontani,Ota05}

While this vertical field configuration has been widely studied both
theoretically and
experimentally,\cite{palacios95,Oh96,Hu96,Tamura98,rontani99,imamura,Tokura99,bart00,Moreno00,Burkard00,amaha,pi,Ono,rontani2,Partoens,Pi2,austing2,Rontani02,Barberan02,Yang02,Barberan03,Park03,ancilotto,rontani,austing,Jacob04,Park,Chwiej}
few works have been devoted to the \emph{in-plane} field
configuration,\cite{Tokura00,Burkard00,Sanchez01,ancilotto,devis2}
which is more complicated to simulate and less intuitive,
principally because of the lack of the cylindrical symmetry and of
analytical solutions for the SP states. Furthermore, since in this
configuration angular momentum is not a good quantum number, the
computational effort increases very rapidly with the electron number
(see Sec.~II B).

On the other hand, in vertically coupled QDs an in-plane magnetic
field $B_\parallel$ modulates the tunneling energy, and one may hope
to induce transitions between few-particle states, reminiscent of
those that one can obtain by varying the tunneling energy via the
barrier potential.\cite{rontani99} In this way, different coupling
regimes would be investigated within just one sample, which is of
course an advantage, since it is almost impossible to growth samples
completely identical except for the barrier width. It should be
stressed that when the tunneling is modulated by an in-plane
magnetic field in vertically coupled QDs, the other energy scales,
in particular the Coulomb interaction, are only weakly affected by
the field. Due to its importance to the QD-based implementation of
quantum-information processing,\cite{Burkard00} such field
configuration has been already investigated in the two-electron
case,\cite{devis2} whereas less attention has been devoted to the
case where more than two carriers are present, in particular when
$B$ has both an in-plane and a perpendicular component, which is
relevant for the transport spectroscopy of such samples. In this
case we expect an interesting physics: in fact both in-plane (due to
$B_\perp$) and vertical (due to $B_\parallel$) correlations are
involved in the transitions between quantum states.

In this paper we theoretically investigate the stability of
field-induced quantum states of few electrons in AMs. The main focus
is on correlation effects, largely enhanced by the field, which
determine different mechanisms (which may interfere between each
other) driving the transition between different ground states,
depending on the field direction and inter-dot coupling. In our
analysis we consider three realistic samples with different
inter-dot barrier, with the aim of investigating the phenomenology
occurring between the two opposite limits of strong coupling and
molecular dissociation. Our numerical approach is based on a
real-space description of SP states, which takes into account
the complexity of real samples. Because of the competition between
kinetic energy, Coulomb interaction, and Zeeman energy, it is often
necessary to treat exactly (in contrast, e.g, to mean-field methods)
the few-body Hamiltonian of this kind of samples. Our method of
choice to include carrier-carrier Coulomb interaction is the Full
Configuration Interaction (FCI) approach,\cite{rontani3} which
proved to be accurate and
reliable,\cite{rontani,Ota05,Garcia05,Bertoni05} although limited to
few electrons. In order to stress correlation effects, we contrast
our numerical results with SP and Hartree-Fock predictions of the
stable quantum phases. We will see that correlation effects play a
crucial role in determining the electronic properties of the system,
and that schematic pictures neglecting these phenomena do not lead
to correct predictions.

The paper is organized as follows: In Sec.~II we illustrate our
numerical approach to the calculation of SP and interacting states;
in Sec.~III we discuss the results concerning different samples in a
magnetic field of arbitrary direction; in Sec. IV we summarize our
findings.

\section{THEORETICAL MODEL}

We consider $N$ interacting electrons in an AM structure. Since
we are interested in weakly confined devices, the energy region of our concern is relatively close to the
semiconductor band gap and we may use the single-band approximation for
the conduction band. In this framework
the carriers are described by the effective-mass Hamiltonian
\begin{eqnarray}\label{hmolti}
H =\sum_{i=1}^{N}\Bigg{[}\frac{1}{2m^{*}}
\Big{(}-i\hbar\nabla_{i}+\frac{|e|}{c}\mathbf{A}(\mathbf{r}_{i})\Big{)}^{2}
+V(\mathbf{r}_{i})\Bigg{]}  \nonumber \\
+\frac{1}{2}\sum_{i\neq j}\frac{e^{2}}{\kappa|
\mathbf{r}_{i}-\mathbf{r}_{j}|}+\frac{g^{*}\mu_{\mathrm{B}}}{\hbar}\mathbf{B}\cdot\mathbf{S}.
\end{eqnarray}
Here $m^{*}$, $\kappa$, and $g^{*}$ are the effective mass,
dielectric constant, and $g$-factor, respectively, $\mu_\mathrm{B}$ is
the Bohr magneton, $\mathbf{S}$ is the total spin,
$\mathbf{A}(\mathbf{r})$ is the vector potential at position
$\mathbf{r}$, and $\mathbf{B}$ is the magnetic field, with
$\mathbf{B}=\nabla\times\mathbf{A}$. Equation (\ref{hmolti})
neglects non-parabolicity effects, but it includes otherwise in the
confinement potential $V(\mathbf{r})$ the full three-dimensional
nature of quantum states in realistic samples, such as layer width,
tunneling, and finite band offsets.

In the following the potential $V(\mathbf{r})$ describes two
identical vertically coupled disk-shaped QDs. Since in this kind of
samples the confinement is much tighter in the growth direction $z$
than in the $xy$ plane, we separate the potential as
$V(\mathbf{r})=V(x,y)+V(z)$, where $V(z)$ represents two identical
square quantum wells of width $L_{W}$, separated by a barrier of
width $L_{b}$ and conduction band mismatch $V_{0}$. We perform the
usual choice of a parabolic in-plane confinement of natural
frequency $\omega_0$:
\begin{equation}\label{parabolic}
V(x,y)=\frac{1}{2}m^{*}\omega_{0}^{2}(x^{2}+y^{2}).
\end{equation}
Adequacy of Eq.~(\ref{parabolic}) has been demonstrated by both
theoretical calculations\cite{kumar,stopa} and far infra-red
spectroscopy experiments.\cite{heitmann,broido} Note, however, that
our numerical approach does not assume any symmetry and may treat
different confinement potentials; in particular, the vector
potential $\mathbf{A}(\mathbf{r})$ is not limited to describe
$z$-directed field. We suppose that the active region of the AM is
sufficiently well separated from the top and substrate contacts that
we can reasonably neglect interactions with the leads in our
model.\cite{rontani} The energy splitting $\hbar\omega_{0}$ between
SP eigenvalues of the in-plane confinement,\cite{jacak} given by
Eq.~(\ref{parabolic}), is an input parameter of the model. Such
quantity  mimics the effects of electrode screening and interaction
with the environment, and in real charging experiments it is
modified by the Schottky gate.\cite{rontani} Here, for the sake of
simplicity, $\hbar\omega_{0}$ is kept fixed as $N$ is varied. A
qualitative discussion of this issue is postponed to the end of
Sec.~\ref{s:results}.

Schematically, our algorithm is the following. First, we numerically
calculate the electron SP states, mapping the SP Hamiltonian on a
real-space grid. Then, the SP orbitals thus obtained are used to
evaluate Coulomb matrix elements and finally to represent the
interacting Hamiltonian on the basis of Slater determinants (SDs),
according to the FCI approach. These steps are described below in
detail.

\subsection{Single-particle states}
\label{sub:Single--states}

The SP energies $\varepsilon_{\alpha}$ and wave functions
$\phi_\alpha(\textbf{r})$ are obtained from the numerical solution
of the eigenvalue problem associated to the Hamiltonian
\begin{equation}\label{single}
H_{0}(\mathbf{r})=\frac{1}{2m^{*}}
\Big{(}-i\hbar\nabla+\frac{|e|}{c}\mathbf{A}(\mathbf{r})\Big{)}^{2}+V(\mathbf{r}),
\end{equation}
which is the SP term appearing in the first line of
Eq.~(\ref{hmolti}). We include in Eq.~(\ref{single}) a magnetic
field of arbitrary direction via the vector potential
$\mathbf{A}=\frac{1}{2}\left(\mathbf{B}\times\mathbf{r}\right)$
(symmetric gauge). Specifically, we find the eigenstates of
Eq.~(\ref{single}) by mapping it on a real-space grid of
$N_{\mathrm{grid}}=\prod_{i=1}^3 N_i$ points, identified by the grid
vectors
$\mathbf{r}_{i}=\sum_{k=1}^{3}(\lambda_{i}^{k}-N_{k}/2)\Delta_{k}\mathbf{\hat{e}}_{k}$,
with $\lambda_{i}^{k}=1,\ldots , N_{k}$ and
$\mathbf{\hat{e}}_{1,2,3}=\mathbf{\hat{x}},\mathbf{\hat{y}},\mathbf{\hat{z}}$.
The resulting finite-difference equation can be rewritten in terms
of a discrete eigenvalue problem, as described in detail
elsewhere.\cite{devis1} This results in the diagonalization
of a large, sparse matrix which is performed by the Lanczos
method.\cite{rontani3}

\subsection{Few-particle states}
\label{sec:few-particle states}

Once SP states are obtained, we proceed including the Coulomb
interactions between charge carriers. The full many-body Hamiltonian
$H$ can be rephrased, by means of the second-quantization
formalism,\cite{gross} as
\begin{equation}\label{full-hamiltonian}
    \hat{H} = \hat{H}_0 + \hat{H}_C + \hat{H}_{\mathrm{Z}},
\end{equation}
where $\hat{H}_0$ is the single
particle Hamiltonian
\begin{equation}\label{h0}
\hat{H_{0}}=\sum_\alpha^{N_{\mathrm{SP}}}\sum_\sigma\varepsilon_{\alpha}\hat{c}^{\dagger}_{\alpha
\sigma} \hat{c}_{\alpha \sigma},
\end{equation}
$\hat{H}_C$ is the Coulomb term
\begin{equation}
\hat{H_{C}}=\frac{1}{2}\sum_{\alpha \beta \gamma
\delta}^{N_{\mathrm{SP}}}\sum_{\sigma \sigma^\prime} V_{\alpha \beta
\gamma \delta} \hat{c}^{\dagger}_{\alpha \sigma}
\hat{c}^{\dagger}_{\beta \sigma^\prime} \hat{c}_{\gamma
\sigma^\prime}\hat{c}_{\delta \sigma},
\end{equation}
and $\hat{H}_{\mathrm{Z}}$ is the Zeeman Hamiltonian
\begin{equation}
\hat{H}_{\mathrm{Z}}
=\frac{\mu_{\mathrm{B}}g^{*}}{\hbar}\mathbf{B}\cdot\hat{\mathbf{S}}.
\end{equation}
Here, $\hat{c}^{\dagger}_{\alpha\sigma}$ ($\hat{c}_{\alpha\sigma}$)
creates (destroys) an electron in the spin-orbital
$\phi_{\alpha\sigma}(\mathbf{r},s)
=\phi_{\alpha}(\mathbf{r})\chi_{\sigma}(s)$, where
$\chi_{\sigma}(s)$ is the spinor wave function, and
$\hat{\mathbf{S}}$ is the total spin vector. In the above equations,
$N_{\mathrm{SP}}$ is the number of SP states that are taken into
account.

The Coulomb matrix elements, given by
\begin{equation}\label{Vmatrices}
V_{\alpha \beta \gamma \delta}= \int\phi^{*}_{\alpha}(\mathbf{r})
\phi^{*}_{\beta}(\mathbf{r^\prime}) \frac{e^{2}}
{\kappa|\mathbf{r}-\mathbf{r^\prime}|}
\phi_{\gamma}(\mathbf{r^\prime})
\phi_{\delta}(\mathbf{r})\,\mathrm{d}\mathbf{r}\,\mathrm{d}\mathbf{r^\prime},
\end{equation}
are calculated by numerically integrating the following expression
\begin{equation}
V_{\alpha\beta\gamma\delta}=\frac{e^{2}}{\kappa}\int\mathcal{F}^{-1}\!\Big{[}
\frac{1}{k^{2}}\tilde{\Phi}_{\beta\gamma}(\mathbf{k})\Big{]}\Phi_{\alpha\delta}
(\mathbf{r})\,\mathrm{d}\mathbf{r},
\end{equation}
where
$\Phi_{\alpha\beta}(\mathbf{r})=\phi_{\alpha}^{*}(\mathbf{r})\phi_{\beta}(\mathbf{r})$,
and
$\tilde{\Phi}_{\alpha\beta}(\mathbf{k})=\mathcal{F}[\Phi_{\alpha\beta}(\mathbf{r})]$
is its Fourier transform.

Since typical SP energy spacings computed from the Schr\"odinger
equation associated with Eq.~(\ref{single}) are in the meV range,
while characteristic values of the Zeeman term
$\hat{H}_{\mathrm{Z}}$ are typically two orders of magnitude smaller
(e.g., $\mu_{\mathrm{B}}B=5.79\times 10^{-2}$ meV at 1
T), the effect of $\hat{H}_{\mathrm{Z}}$ is first neglected in the
calculation of the electronic ground states (GSs). Then, the effect
of $\hat{H}_{\mathrm{Z}}$, which just lifts the ($2S+1$)-fold
degeneracy of the total spin $S$, is included as a perturbation to
the first order in the field.

\begin{figure}
\centering
\includegraphics[clip,angle=0,width=0.5\textwidth]{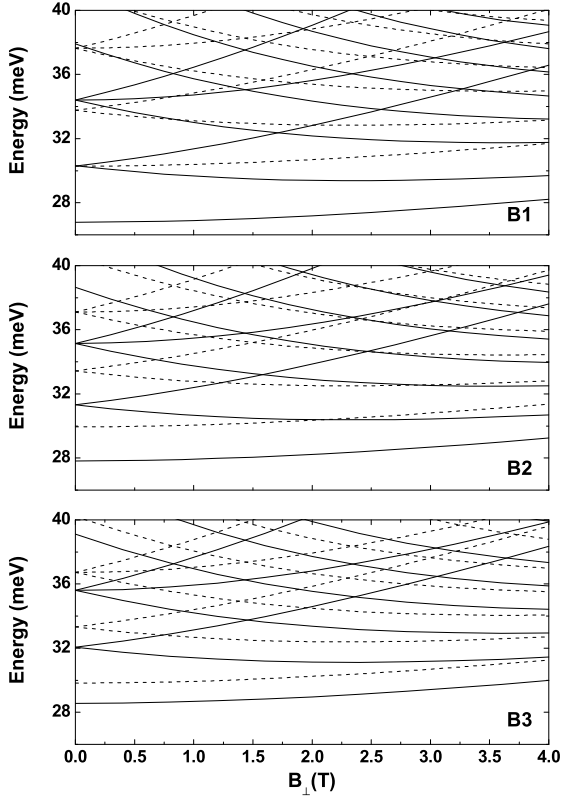}
\caption{Fock-Darwin energy levels of samples \bone, \btwo, \bthree\
(see text) vs.~vertical magnetic field, $B_{\perp}$. Solid and dashed
lines represent energies $\varepsilon_{nm}$ arising from the lowest
S and AS levels of the double quantum well confinement along the growth
direction, respectively.} \label{fig1}
\end{figure}

Then, our algorithm proceeds forming a
basis of SDs, filling $N_{\mathrm{SP}}$ spin-orbitals
$\phi_{\alpha\sigma}(\mathbf{r},s)$, calculated numerically as
described in Sec.~II A, with $N$ electrons in all possible ways.
We assume that the Fock
space generated by the SDs is approximately complete, namely a
generic eigenstate of the many-body Hamiltonian $\hat{H}$ can be
expressed as a linear combination of SDs. One may reduce the
computational effort by exploiting the symmetries of the system. In
fact, by using a suitable combination of the SDs which diagonalize
symmetry-related operators, the Hamiltonian can be written in a
block-diagonal form.\cite{rontani3} The larger the symmetry of the
system, the smaller the dimensions of the blocks and the time
necessary to complete the full diagonalization. In the present case,
an in-plane component of the magnetic field removes the cylindrical
symmetry $C_{\infty h}$, which is often exploited in the vertical
field arrangement of cylindrically symmetric QDs. However, the
Hamiltonian still commutes with the square total spin $\hat{S}^{2}$
and $z$-component $\hat{S}_{z}$ operators, respectively (neglecting
$\hat{H}_{\mathrm{Z}}$). Therefore, the subspaces are appropriately
labeled by the values of $S$ and the minimum positive value of
$S_{z}$ consistent with $S$. Once the effective subspace is
selected, our code performs a unitary transformation in order to
rewrite $\hat{H}$ in a block-diagonal form. Finally, matrices
obtained in this way are handled by a Lanczos routine included in
the package {\tt donrodrigo}\cite{donrodrigo} run on a SP3 IBM
system.

\section{Results}\label{s:results}

\begin{figure}
\centering
\includegraphics[clip,angle=0,width=0.5\textwidth]{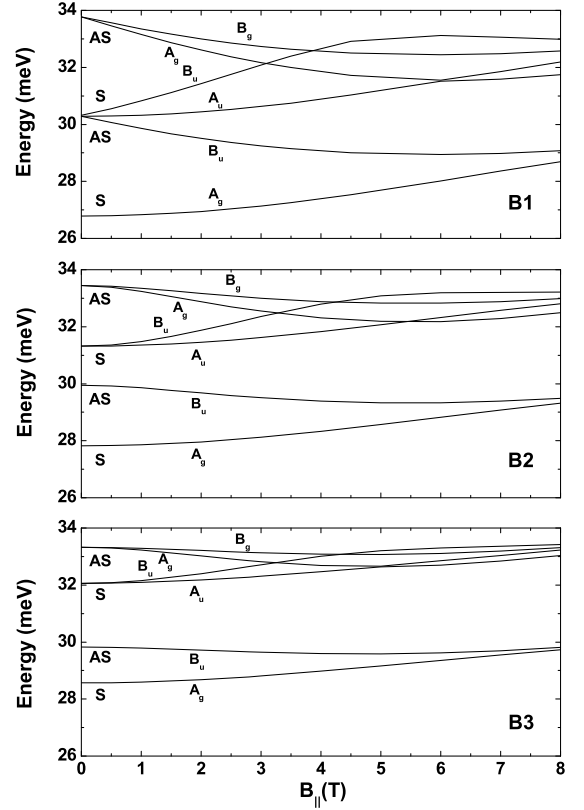}
\caption{Single-particle states of samples \bone, \btwo, \bthree\
vs.~in-plane field $B_{\parallel}$ at $B_\perp=0$.} \label{fig2}
\end{figure}

\begin{figure}
\centering
\includegraphics[clip,angle=0,width=0.5\textwidth]{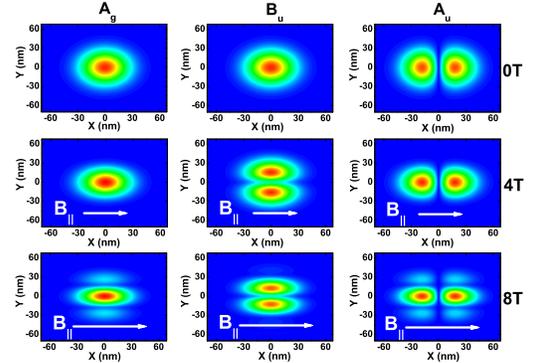}
\caption{(Color online) Contour plots of $|\phi(x,y)|^{2}$, at a
fixed value of $z$ inside one well, of selected SP states (see
labels on top) of sample \bthree\ with an in-plane field
$B_\parallel=0$ (top row), $B_\parallel=4$ T (center row) and
$B_\parallel=8$ T (bottom row). $B_{\parallel}$ lies along the $x$
axis. Symmetry labels refer to the lowest-energy SP orbitals of a
given representation (cf.~bottom panel of
Fig.~\ref{fig2}).}\label{fig3}
\end{figure}

In this section we present the calculated field-dependent GSs of
AMs, covering the regimes from strong to weak coupling. The
application of the FCI method ensures that our results have
comparable accuracy in different regimes. To this end, we have
selected three realistic samples based on a AlGaAs/GaAs
heterostructure, labeled \bone, \btwo, \bthree, with inter-dot
barriers $L_{b} = 2.5$, 3.2, and 4.0 nm, respectively. The sample
with intermediate barrier shows a full molecular
character, while \bone\ and \bthree\ represent the sigle QD and the
molecular dissociation limits, respectively.

Except for the barrier width, parameters are common to all samples.
The lateral confinement is $\hbar\omega_{0}=3.5$ meV, the
band-offset between barrier and well materials of the
heterostructure is $V_{0}=300$ meV, each well width of the double
quantum well potential is $L_W = 12$ nm. Material parameters
appropriate to GaAs are $m^*/m_e=0.067$, $\kappa=12.9$, $g^*=-0.44$.
All parameters refer to a set of samples\cite{amaha,rontani}
described in literature and processed at the University of Tokyo.

\subsection{Single-particle states}
\label{sec:sp_states}

SP states in an arbitrary field have been obtained according to
Sec.~\ref{sub:Single--states} with $N_{1}=N_{2}=80$,
$N_{3}=128$; the resulting $819200\times 819200$ sparse Hamiltonian
matrix is diagonalized by means of the Lanczos
algorithm. The grid parameters have been chosen in order to reproduce the
lowest analytical zero field (Fock-Darwin) energies $\varepsilon_{nm}$
with an accuracy better then ~0.5 meV.

Before presenting our results for specific samples, let us summarize
the properties of SP states in a single QD with parabolic in-plane
confinement and vertical magnetic field $B_\perp$. In this case SP
states are analytical and given by the so-called Fock-Darwin states
(FD), with energies
$\varepsilon_{nm}=\hbar\Omega(2n+|m|+1)-(\hbar\omega_{c}/2)m$,
$n=0,1,2,\ldots$ and $m=0,\pm1,\pm2,\ldots$ being the principal and
azimuthal quantum numbers, respectively. The effective oscillator
frequency $\Omega$ is defined by
$\Omega=\sqrt{\omega_{0}^{2}+\omega_{c}^{2}/4}$, and
$\omega_{c}=eB/m^{*}c$ is the cyclotron frequency. A perpendicular
magnetic field splits the degeneracies of the states and reduces the
energy gap between energy levels with increasing angular momentum
(see solid lines in Fig.~\ref{fig1}); this favors transitions in the
few-electron states in order to increase in-plane correlations, as
explained in the next section. In symmetric AMs, FD levels are
replicated, rigidly shifted by $\Delta_{\mathrm{SAS}}$, the energy
gap between the lowest symmetric (S) and anti-symmetric (AS) states
arising from the double-well potential along the growth direction
(highest confined states along the growth direction need not be
considered if $L_W$ is sufficiently small).\cite{rontani} Figure
\ref{fig1} shows the S (solid lines) and AS (dashed lines) replicas
of FD states for the three samples as a function of $B_\perp$. Note
that the S/AS labeling is valid only as far as the field is entirely
along the $z$ axis, namely the motions in the $xy$ plane and along
$z$ are completely uncoupled, and the splitting
$\Delta_{\mathrm{SAS}}$ does not depend on $B_{\perp}$. The symmetry
group, in this highly symmetric case, is $C_{\infty h}$.

\begin{figure}
\centering
\includegraphics[clip,angle=0,width=0.45\textwidth]{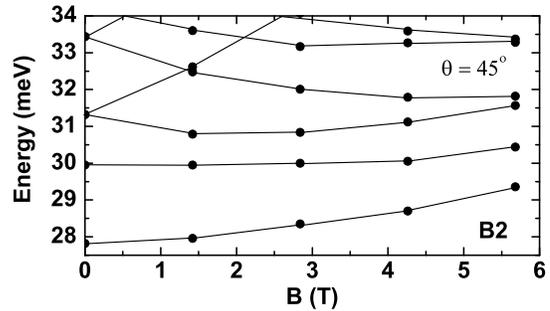}
\caption{Calculated single-particle energy levels of sample \btwo\
vs.~field, $\textbf{B}=B_{\parallel}\hat{\mathbf{x}}+B_{\perp}\hat{\mathbf{z}}$,
forming a 45$^{\circ}$ angle with respect
to the growth direction of the heterostructure.} \label{fig4}
\end{figure}

Figure \ref{fig2} shows calculated SP levels for \bone, \btwo,
\bthree\ as a function of the in-plane field $B_\parallel$. Contrary
to the vertical-field configuration, the parallel field
$B_\parallel$, which we take to lie along the $x$ axis, couples the
motion along the in-plane $y$ and the vertical $z$ directions,
and it reduces the symmetry group of the system from
$C_{\infty h}$ to $C_{2h}$. Therefore, SP wave functions lose their
well defined component of the orbital angular momentum as well as
the S/AS character along $z$. Besides, the states labeled S and AS
when $B_{\parallel}=0$, respectively, get closer as $B_\parallel$
increases, and tunneling is progressively suppressed. Clearly, the
tunneling suppression occurs at different values of the field for
different samples, since the in-plane field may significantly affect
tunneling only when $\omega_{c}^\parallel=eB_\parallel/m^{*}c\sim
\Delta_{\mathrm{SAS}}/\hbar$. \cite{devis2}

Although the application of $B_\parallel$ strongly reduces the
symmetry group of the system, it still preserves two residual
symmetries: i) the reflection $x\rightarrow -x$ with respect to the
$yz$ plane perpendicular to the field direction ii) the rotation of
180$^{\circ}$ around the $x$ axis. The four irreducible
representations of the residual group $C_{2h}$, namely $A_{u}$,
$A_{g}$, $B_{u}$, and $B_{g}$,\cite{landau} are associated to the
appropriate SP states in Fig.~\ref{fig2}, together with the
alternative labeling at zero field. It is possible to understand
qualitatively the behavior of the SP states with the parallel field
$B_{\parallel}$ by taking into account the energy gap between the
orbitals with the same symmetry; the smaller the gap, the larger the
``repulsion'' between levels. For example, we expect strong
repulsion between the second and the fourth state, and a weaker
effect for the first and the fifth state because of the larger
energy gap between them.

The effect of the in-plane field on SP states is demonstrated in
Fig.~\ref{fig3}, which shows the SP charge densities
$|\phi(x,y)|^{2}$ at a fixed value of $z$ inside one well for the
three lowest levels (their quantum numbers are indicated in
the figure), for three selected values of $B_\parallel$. The field
increases the confinement, squeezing the electronic wave functions.
Moreover, FD orbitals with different angular momentum and
belonging to different energy shells are mixed by $B_{\parallel}$,
leading to charge density modulation and nodal surfaces.

In presence of a magnetic field of arbitrary direction (tilted
magnetic field) also the $C_{2h}$ symmetry is broken, and we observe
simultaneously the reduction of both tunneling and intradot energy gaps
between FD states as the field is increased.\cite{devis2} An example
of the SP states behavior in a magnetic field which is tilted at
$\theta=45^{\circ}$ with respect to the growth direction is reported
in Fig.~\ref{fig4}.

\subsection{Few-particle states}

\begin{figure}
\centering
\includegraphics[clip,angle=0,width=0.5\textwidth]{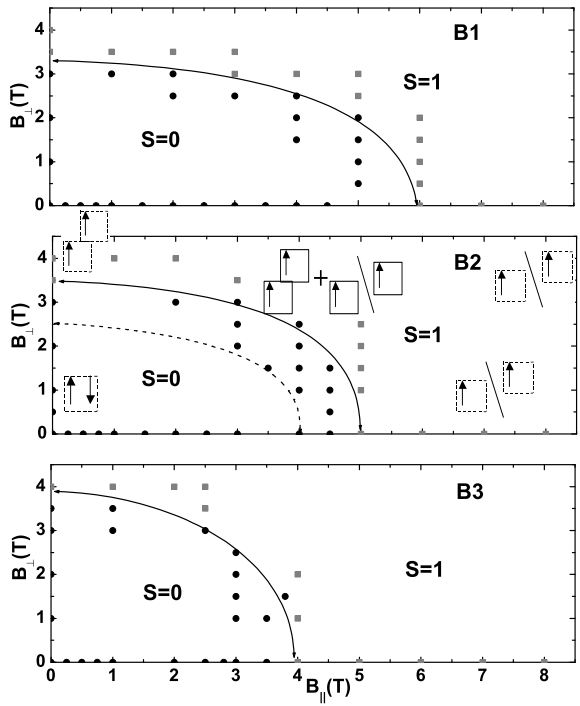}
\caption{$N=2$ stability diagram in the
$(B_{\parallel},B_{\perp})$ plane for \bone, \btwo, and \bthree,
as predicted by the FCI calculations.
Gray squares: $S=1$ GS; black circles: $S=0$ GS. Solid
lines are guides to the eye. The dashed line in \btwo\ indicates
the predicted SSD singlet/triplet boundary. Insets: main components
of the ground state wave functions (see text).}\label{fig5}
\end{figure}

In the following, we will investigate the stability of few-particles
($2\le N\le 5$) quantum phases as a function of an applied field
with both in-plane and vertical components. In AMs, due to tunneling
and the ensuing formation of bonding and anti-bonding levels, the
number of SP orbitals which ensures convergence of the FCI
calculation depends on the splitting, $\Delta_{\mathrm{SAS}}$, and
on the number of charge carriers, $N$, we take into account. Results
presented in this work are obtained using $N_{\mathrm{SP}}=20$ for
$2 \leq N \leq 4$. In order to limit the computational effort, which increases
very rapidly with the number of electrons, we reduce to
$N_{\mathrm{SP}}=15$ the set of SP states for $N=5$.
With these values, the calculated SP states, which at arbitrary values of the field
are in general non degenerate, turn continuously into close
zero field degenerate shells. The convergence of the FCI calculations at zero field
with respect to the selected SP states has been carefully investigated
in Ref.~\onlinecite{rontani3}.

In order to stress the role of correlations, we will contrast the
exact correlated states, obtained by the FCI method, with i) the
quantum phases predicted in a SP picture, i.e., filling SP states
according to the \emph{Aufbau} principle, and ii) a Hartree-Fock-like
approach.

In the SP picture the few-electron non-interacting GS energy is
\begin{equation}
E_{SP}=\sum_{\alpha}\sum_{\sigma}\varepsilon_{\alpha}n_{\alpha\sigma},
\end{equation}
where $\varepsilon_{\alpha}$ is the SP energy of the $\alpha$-th
orbital and $n_{\alpha\sigma}$ is the occupation number of the
$\alpha$-th SP orbital with spin $\sigma$. The second model is
reminiscent of the Hartree-Fock approach: In this approximation, we
calculate the energy of an electronic configuration as the
expectation value of the interacting Hamiltonian on the most
weighted SD singled out from the FCI expansion of the correlated
wave function. We call this single SD (SSD) approximation. The SSD
energy is given by
\begin{eqnarray}
&&E_{SSD}=\sum_{\alpha}\sum_{\sigma}\varepsilon_{\alpha}n_{\alpha\sigma}+\frac{1}{2}
\sum_{\alpha\ne
\beta}U_{\alpha\beta}\sum_{\sigma\sigma^\prime}n_{\alpha\sigma}n_{\beta\sigma^\prime}
\nonumber\\ &&\quad -\frac{1}{2}\sum_{\alpha\ne
\beta}K_{\alpha\beta}\sum_{\sigma}n_{\alpha\sigma}n_{\beta\sigma}
+\sum_{\alpha}U_{\alpha\alpha}n_{\alpha\uparrow}n_{\alpha\downarrow}.
\end{eqnarray}
Here the interaction between charge carriers is described by the
integrals
\begin{eqnarray}
U_{\alpha\beta}&=&\int \!\!\!
\int|\phi_{\alpha}(\mathbf{r_{1}})|^{2}\frac{e^{2}}{\kappa|\mathbf{r_{1}}-\mathbf{r_{2}}|
}|\phi_{\beta}(\mathbf{r_{2}})|^{2}\,\mathrm{d}\mathbf{r}_1\,\mathrm{d}\mathbf{r}_2, \label{eq:U}\\
K_{\alpha\beta}&=&\int \!\!\!
\int\phi_{\alpha}^{*}(\mathbf{r_{1}})\phi_{\beta}^{*}(\mathbf{r_{2}})
\frac{e^{2}}{\kappa|\mathbf{r_{1}}-\mathbf{r_{2}}|}\times
\\ \nonumber
&\,\,\times&\phi_{\alpha}(\mathbf{r_{2}})\phi_{\beta}(\mathbf{r_{1}
})\,\mathrm{d}\mathbf{r_{1}}\,\mathrm{d}\mathbf{r_{2}},
\label{eq:UK}
\end{eqnarray}
where $U_{\alpha\beta}$ is the direct Coulomb integral, accounting
for the repulsion between two electrons occupying orbitals $\alpha$
and $\beta$, and $K_{\alpha\beta}$ is the exchange integral, giving
the exchange interaction between electrons with parallel spins.
$U_{\alpha\beta}$ and $K_{\alpha\beta}$ correspond to
$V_{\alpha\beta\beta\alpha}$ and $V_{\alpha\beta\alpha\beta}$ in
Eq.~(\ref{Vmatrices}), respectively. Note that in both SP and SSD
models the GS is given by just one configuration, while, in general,
the true GS is a linear combination of different electronic SDs, due
to correlation. We also neglect the Zeeman coupling to the field
at this level.

Below we shall discuss the GS of few electrons as a function of
the magnetic field intensity and direction. In several cases, we
will indicate schematically (see following figures) the GS
configuration, as predicted either by the FCI method or by the SSD
calculation, in terms of arrows, pointing either upwards or
downwards, filling in either left or right boxes (with respect to a
diagonal line). The arrows stand for electron spin,
while superposed boxes indicate SP orbitals of increasing energy. In
the vertical-field configuration, left and right boxes indicate S
and AS levels, respectively (right boxes are not shown if no AS
states are occupied). If a finite $B_{\parallel}$ component is
present, the boxes indicate the orbitals which evolve in a
continuous manner from those at $B_{\parallel}=0$ as $B_{\parallel}$
is switched on (see Fig.~\ref{fig2}). For example, the lowest-energy
S (AS) $s$ and $p$ levels at $B_{\parallel}=0$ evolve continuously
into $A_g$ and $A_u$ ($B_u$ and $A_g$) levels, respectively, as
$B_{\parallel}$ increases, while remaining identified pictorially by
the same boxes. For FCI calculation, we show only the most weighted
configurations. In some cases FCI and SSD methods predict the same
stable phase, in which case we use dotted boxes. When the two
methods disagree, we use solid boxes and indicate the exact (FCI)
result only. Typically, for a given sample and number of carriers,
the set of stable GSs predicted by the SSD method is a subset of the
FCI stable phases.

In all stability diagrams shown below, calculated GSs are indicated with
dots, and lines are only a guide to the eye through the calculated
points. Due to numerical limitations, mainly related to the lack of
cylindrical symmetry which requires numerical calculation of the SP
states, we have limited our calculations to a relatively coarse grid
of points in the $(B_\perp,B_\parallel)$ plane. Although we believe
that we have determined most of the GSs which are stable in the
different regimes, it is possible that we miss some phases which are
stable in a small range of fields.

We first consider the two-electron case.\cite{devis2} In a SP
picture, the lowest SP state is doubly filled in a singlet ($S=0$)
configuration which is stable in all field regimes, since no level
crossing occurs in the lowest SP state (see Figs.~\ref{fig1}
and \ref{fig2}). Figure \ref{fig5} shows, instead, that FCI predicts
a singlet/triplet transition at finite fields, both along the
$B_{\perp}$ and $B_{\parallel}$ axes. For
\btwo\ the SSD phase boundary is also reported in Fig.~\ref{fig5}
(no qualitative differences are expected for the other samples).

Let us focus for the moment on the SSD stability diagram of
Fig.~\ref{fig5}. At low magnetic field the ground state has a
singlet character, with two electrons sitting on the same orbital.
As the field is increased, we observe a singlet/triplet transition,
which is different in character depending whether $B_\perp$ or
$B_\parallel$ is varied.\cite{devis2} In the former case, only S
levels $s$ and $p$ (equivalent to $A_g$, $A_u$ levels, respectively)
are involved: the triplet configuration minimizes the Coulomb
interaction by promoting an electron from a $s$ to a $p$ orbital,
which is more spatially delocalized, i.e., $U_{ss} > U_{sp}$. In
this sense, only in-plane degrees of freedom play a role in the
$B_\perp$-induced transition. The ``Hartree'' energy gain
$U_{sp}-U_{ss}$ compensates the cost in term of SP energy due to the
$s\rightarrow p$ orbital promotion. Moreover, the exchange
interaction energy gain, given by the exchange integral $K_{sp}$,
favors the ``ferromagnetic'' triplet configuration ($S=1$) with
respect to the ``antiferromagnetic'' singlet, when the two electrons
sit on the $s$ and $p$ orbitals, respectively. Therefore, the
electrostatic energy of the triplet configuration $\approx
U_{sp}-K_{sp}$ is further reduced with respect to the singlet one,
$\approx U_{sp}$.

In the in-plane configuration, instead, the tunneling energy
separating the lowest-energy $A_g$ and $B_u$ orbitals is very
effectively reduced by $B_\parallel$ (see Fig.~\ref{fig2}). Both
orbitals are therefore involved in the formation of singlet and
triplet states. In the SSD approximation singlet and triplet states,
the former made of a doubly occupied $A_g$ orbital and the latter of
two parallel-spin electrons sitting on the $A_g$ and $B_u$ orbitals,
respectively, tend to the same orbital energy. However, the small
exchange interaction $K_{A_gB_u}$ and the Zeeman term of the
$S=1$ state, both included in the SSD calculation,
weakly favor the triplet when the tunneling energy is reduced, above
a critical value of $B_\parallel$.

Although the SSD scheme is able to describe, on a qualitative level,
the singlet/triplet stability regions, neglecting correlation
effects in the SSD scheme strongly underestimates the stability
region of the singlet state, favoring the polarized phase: in fact,
the description of the singlet state in terms of two electrons
sitting on a $A_g$ orbital is poor especially at high field, where
an increasing contribution from the $B_u$ orbital gives rise to a
spatial correlation along the growth direction, i.e., with the two
electrons sitting on opposite QDs, as we have shown in a previous
work.\cite{devis2,note} The vertical correlation lowers the energy
of the singlet state, making it stable in a larger
$B_\parallel$-region, although the Zeeman coupling favors the
triplet configuration. Indeed, the stability of the singlet state in
FCI may be interpreted as resulting from a ``superexchange''
effect.\cite{devistesi} The FCI singlet configuration can be better
understood after a unitary transformation of the orbital basis to
states localized on either one or the other dot. By  schematizing
the AM as a two-site lattice at half filling,
the appropriate conceptual framework, when the QDs
are weakly coupled, is that of the Hubbard
model,\cite{hubbard,gutzwiller} where the system has a tendency to
``antiferromagnetism'', since two electrons in a singlet state can
lower their energy by virtually hopping from one to the other dot,
while the same process is prohibited for a triplet state by Pauli
blocking. When tunneling energy does not pay off for antiparallel
alignment, the triplet phase becomes favorable.

On the other hand, a SSD scheme would neglect also the correlation
regarding the triplet state when both $B_\perp$ and $B_\parallel$
are present, namely the central region in the diagrams of
Fig.~\ref{fig5}. Here the triplet state is given by the sum of
mainly two components, the first one involving $A_g$ and $A_u$
orbitals, and the second one involving $A_g$ and $B_u$ orbitals,
with coefficients depending on the field strength and direction. In
this case the effects induced by both $B_\parallel$ and $B_\perp$
compete in a non-trivial manner, as well as vertical and in-plane
correlations.\cite{nota-labelling}

Finally, we note that the stability region of the singlet state
\emph{with respect to} $B_\parallel$ depends on the inter-dot
coupling, and it decreases going from \bone\ to \bthree. In fact, as
we have seen before, the in-plane field affects significantly the
tunneling only when $\omega_{c}^\parallel=eB_\parallel/m^{*}c$ is
comparable with $\Delta_{\mathrm{SAS}}/\hbar$.\cite{devis2}
Consistently with the above interpretation of the siglet/triplet
transition, the critical value of $B_\perp$ does not depend on
tunneling strength.

\begin{figure}
\centering
\includegraphics[clip,angle=0,width=0.45\textwidth]{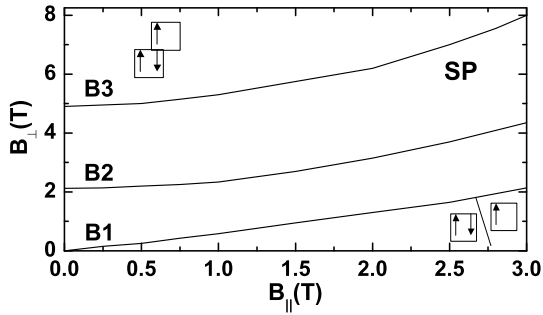}
\caption{$N=3$ stability diagram in the $(B_{\parallel},B_{\perp})$
plane for \bone, \btwo, and \bthree, obtained within the SP non-interacting
scheme.} \label{fig6}
\end{figure}

\begin{figure}
\centering
\includegraphics[clip,angle=0,width=0.5\textwidth]{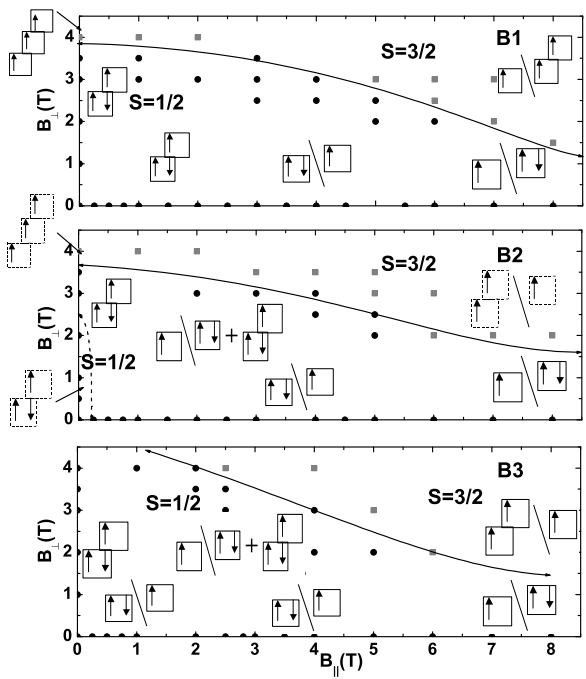}
\caption{$N=3$ stability diagram in the
$(B_{\parallel},B_{\perp})$ plane for \bone, \btwo, and \bthree,
as predicted by the FCI calculations.
Gray squares: $S=3/2$ GS; black circles: $S=1/2$ GS. Solid lines are
guides to the eye. The dashed line for \btwo\ is the
SSD prediction. Insets: main components of the GS wave functions
(see text). } \label{fig7}
\end{figure}

\begin{figure}
\centering
\includegraphics[clip,angle=0,width=0.5\textwidth]{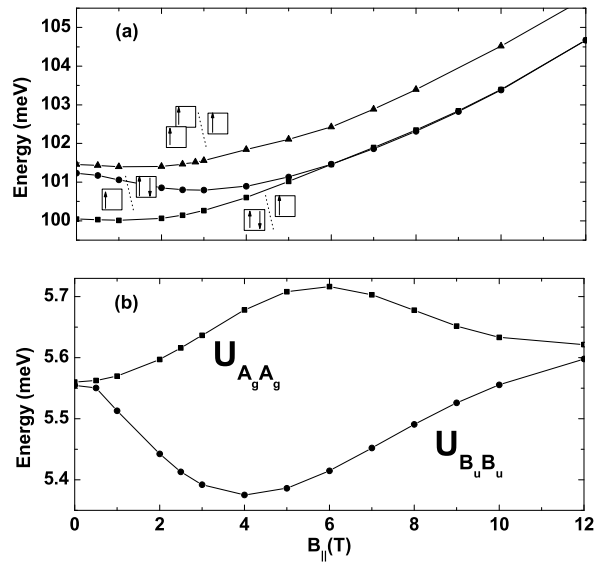}
\caption{(a) Lowest energy levels for $N=3$ vs.~in-plane field
$B_\parallel$ at $B_\perp$=0, for sample \bthree. (b) Coulomb
integrals $U_{A_gA_g}$ and $U_{B_uB_u}$ vs.~in-plane field
$B_\parallel$ at $B_\perp$=0, for the same sample. The Coulomb
integrals $U_{\alpha\alpha}$ are defined in Eq.~(\ref{eq:U}).}
\label{fig8}
\end{figure}

We next consider the $N=3$ case. In the SP non-interacting scheme
(Fig.~\ref{fig6}) only two phases are present, both with $S$=$1/2$.
The field $B_\perp$ at which the two GS energies cross is given by
the intersection of the S $s$ ($\equiv A_g$) and $p$ ($\equiv A_u$)
levels (Fig.~\ref{fig1}). Figure \ref{fig7} shows the stability
diagram obtained from the FCI calculation; for \btwo\ the phase
boundary obtained from the SSD approach is also shown (dashed line).
Although the SSD approximation seems to describe reasonably well the
dynamics along the $B_\perp$ axis, where GS configurations coincide
with the FCI ones, except a slight shift of the phase boundary at
lower fields with respect to FCI, this scheme fails completely at
finite $B_\parallel$. Indeed, SSD severely underestimates the
stability region of the unpolarized phase ($S$=1/2) with respect to
the polarized one ($S$=3/2). This fact is mainly connected to
correlation effects, as we shall see in the following.

Let us stick for a moment to the vertical field configuration. As already
for $N=2$, also for $N=3$ increasing $B_\perp$ determines an
enhancement of the in-plane correlation, leaving unaffected the
motion along the growth direction. For \bone\ and \btwo, electrons
in the S orbitals undergo a number of transitions where $p$ and $d$
levels are successively populated in order to minimize the Coulomb
repulsion. Eventually, the so called maximum density droplet
(MDD),\cite{yang,reimann,ferconi,taruchaMDD} i.e., the densest
spin-polarized configuration possible, is reached. For \bthree\ this
phase is not observed at $B_\parallel=0$ in the range of $B_\perp$
explored (Fig.~\ref{fig7}).

If $B_\parallel$ is small or zero, electrons are prevented from
occupying AS orbitals by the kinetic energy cost
$\Delta_{\mathrm{SAS}}$. As $B_\parallel$ is increased, however,
$\Delta_{\mathrm{SAS}}$ decreases. The range of fields where AS
occupation takes place depends, of course, on the zero field
tunneling energy, as it is easily seen comparing the GS character at
low $B_{\parallel}$ of the \bthree\ sample, where electrons occupy
both S and AS states, with that of \bone\ and \btwo\ samples, where
AS states are not occupied. From this point of view,
\emph{increasing $B_\parallel$ is analogous to decreasing $L_b$}. We
will find a similar behavior for $N=4,5$.

On the other hand, in addition to the reduction of the tunneling
splitting $\Delta_{\mathrm{SAS}}$, a finite value of $B_\parallel$
determines a redistribution of the carrier density, coupling the $y$
and $z$ degrees of freedom, reducing the spatial symmetry, and
inducing orbital hybridization, as discussed in
Sec.~\ref{sec:sp_states}. From this point of view, \emph{increasing
$B_\parallel$ is not equivalent to decreasing $L_b$}, since it
induces GS configurations different from those occurring at large
$L_b$. Moreover, as we show below, the high-$B_{\parallel}$ phases
are brought about by genuine correlation effects, and in fact they
are missed by the SSD approximation.

Let us consider the spin phase at high $B_\parallel$ and low
$B_\perp$ in Fig.~\ref{fig7}, having the $B_u$ orbital doubly
occupied in its most-weighted SD. At high $B_\parallel$, the
orbitals $A_g$, $B_u$ are almost degenerate (see Fig.~\ref{fig2}).
Therefore, we would expect the two configurations with two electrons
either in the $A_g$ or in the $B_u$ level to occur with like
probabilities. However, intra-orbital Coulomb repulsion is lower for
the more delocalized $B_u$ state. This is shown for \bthree\ in
Fig.~\ref{fig8}(b), where we report the behavior of
$U_{\alpha\alpha}$ [defined in Eq.~(\ref{eq:U})], where $\alpha$
takes the value $A_g$ and $B_u$, respectively. The modulation of
$U_{A_gA_g}$ and $U_{B_uB_u}$ with $B_\parallel$ should not come as
a surprise, since the SP orbitals are strongly affected by the field
(Fig.~\ref{fig3}), with the appearance of nodal surfaces even in the
lowest-energy SP orbital $A_g$. These integrals allow to estimate
the value of the Coulomb repulsion between two electrons both
sitting in the $A_g$ or in the $B_u$ orbital, as a function of the
in-plane field $B_{\parallel}$. In Fig.~\ref{fig8}(a) we show the
ground- and lowest excited-state energies for \bthree\
vs.~$B_\parallel$. The larger values of $U_{A_gA_g}$ at intermediate
fields partly explains the occurrence of the SD with two electrons
sitting on the $B_u$ orbital. On the other hand, such GS is not
observed in a SSD picture, i.e., neglecting correlation, as shown in
Fig.~\ref{fig7} (note that the $B_\parallel$ field range is larger
in Fig.~\ref{fig8} than in Fig.~\ref{fig7}). At high values of
$B_{\parallel}$, $U_{A_gA_g}\simeq U_{B_uB_u}$ and the state with
two particles on the $B_u$ level becomes almost degenerate with the
one with two particles on the $A_g$ level [Fig.~\ref{fig8}(a)]. The
results for \bone\ and \btwo\ samples (not shown) are similar.

The reduced Coulomb repulsion between electrons on the $B_u$ orbital
is a key concept to understand the stability of the unpolarized
phase $S=1/2$. Indeed, the SD with the $A_g$ orbital singly occupied
and the $B_u$ orbital doubly occupied is always present in the GS
when applying $B_\parallel$ (even though it is not indicated in the
graphs unless it is the dominating one), together with other
important configurations. The reason is that correlation, namely the
mixing of different SDs, allows the system to lower its energy. At
high $B_{\parallel}$, the SD mentioned above  becomes the dominating
one for all considered samples, although close in energy to the SD
with the $A_g$ orbital doubly occupied. As in other cases, in a
tilted magnetic fields the typical SDs appearing in the GS
expansion, in the central region of the diagrams, represent the
competition between vertical correlation, driven by $B_\parallel$,
and in-plane correlation, driven by $B_\perp$.

\begin{figure}
\centering
\includegraphics[clip,angle=0,width=0.45\textwidth]{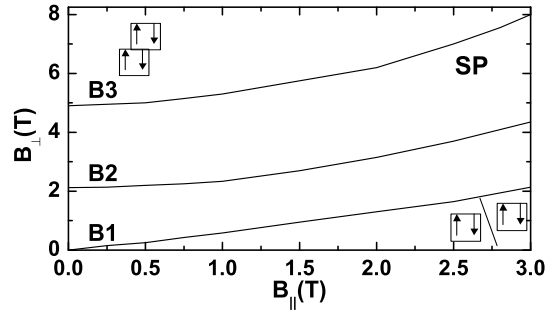}
\caption{$N$=4 stability diagram in the
$(B_{\parallel},B_{\perp})$ plane for \bone, \btwo, and \bthree,
as obtained within the SP non-interacting
scheme.} \label{fig9}
\end{figure}

\begin{figure}
\centering
\includegraphics[clip,angle=0,width=0.45\textwidth]{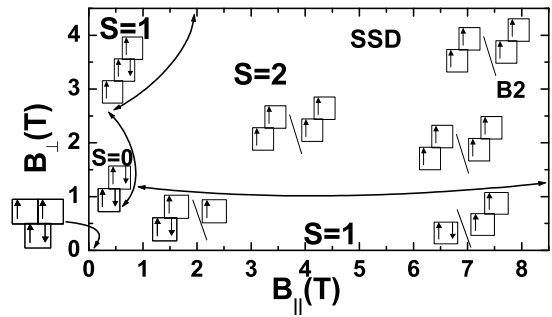}
\caption{$N$=4 stability diagram in the
$(B_{\parallel},B_{\perp})$ plane for \btwo, calculated within the SSD
scheme.} \label{fig10}
\end{figure}

\begin{figure}
\centering
\includegraphics[clip,angle=0,width=0.5\textwidth]{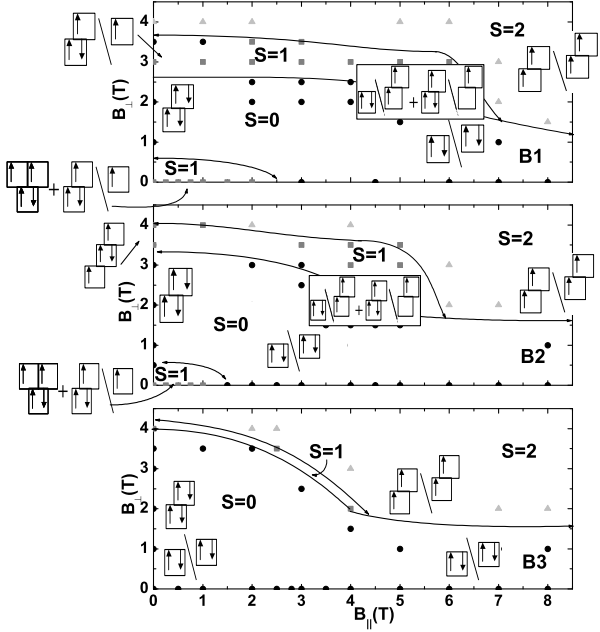}
\caption{$N=4$ stability diagram in the
$(B_{\parallel},B_{\perp})$ plane for \bone, \btwo, and \bthree, as
predicted by the FCI calculations.
Gray triangles: $S=2$ GS; gray squares: $S=1$ GS; black circles: $S=0$ GS.
Solid lines are guides to the eye. Insets: main
components of the GS wave functions (see text).} \label{fig11}
\end{figure}

For $N=4$ the situation is more complicated due to the larger number
of possible phases, and the strong dependence on the inter-dot
coupling strength. SP, SSD, and FCI predictions are presented
separately in Figs.~\ref{fig9}, \ref{fig10}, and \ref{fig11},
respectively. The predictions of the SP scheme in Fig.~\ref{fig9},
if compared with those of FCI of Fig.~\ref{fig11}, are clearly
inadequate to describe the physics of the system: only two
non-interacting singlet phases are stable, differing for the double
occupancy  of either the $A_u$ or the $B_u$ orbital.\cite{nota-s1}

As expected, the FCI calculation (Fig.~\ref{fig11}) predicts the
familiar MDD state formation through
successive occupation of $p$ and $d$ orbitals;
this is also predicted, even though
approximately, by the SSD approximation (Fig.~\ref{fig10}).
On the other hand there are many differences between SSD
(Fig.~\ref{fig10}) and FCI (Fig.~\ref{fig11}) predictions when
$B_\parallel$ is also increased. One major difference is the
enhanced stability of the $S=0$ phase in FCI calculations. This
phase involves the SD made of doubly occupied $A_g$ and $B_u$
orbitals, which is the dominant one in a large field range for all
samples. Analogously to the $N=3$ case, the system is allowed to
lower its energy through occupation of the $B_u$ level as
$B_\parallel$ is increased. The true GS, however, turns out to be a
non trivial linear combination of several other SDs, which suggests
that correlations are dominant here. Indeed, by neglecting
correlation effects, like in the SSD approach, the only possible
energy gain is obtained by aligning the spins in a $S=1$ phase,
leading to the wrong SSD prediction that the $S=1$ phase is always
stable at small $B_\perp$, as shown in Fig.~\ref{fig10}.

Such partially polarized phases, instead, are typically stable only in small
ranges of the field, and their stability also depend on the sample parameters,
as we show below. For $N=4$ the $S=1$ phase at $B\approx 0$ is stable only
in a small field range, and its stability is reduced if the
inter-dot coupling is reduced; in sample \bthree\ this phase is
not stable, no matter how small the field is (Fig.~\ref{fig11}).
Indeed, this phase requires occupation of two S $p$ orbitals, and it
results from the payoff between the gain in exchange energy and the
kinetic energy cost. Therefore, this configuration becomes rapidly
unfavorable with respect to the $S=0$ phase having the $B_u$ orbital
doubly occupied, as $B_\parallel$ increases. In fact, the stability
range of the $S=1$ phase decreases as the zero field tunneling
energy $\Delta_{\mathrm{SAS}}$ decreases with respect to the $s-p$
energy splitting $\hbar\omega_0$. Note that, according to the SSD
data (Fig.~\ref{fig10}), the $S=1$ phase is given by only one SD,
instead of the FCI superposition of two main configurations with
coefficients dependent on the value of the field. This fact, which
indicates the progressive filling of the $B_u$ level while
increasing the field, is neglected by the less accurate SSD picture,
and it is described by SSD model in terms of a transition between
the two SDs mentioned above.
We note in Fig.~\ref{fig11} that a $S=1$ phase sets in also
between the unpolarized ($S=0$) and the completely
polarized ($S=2$) spin case. This phase, characterized by a non
trivial combination of configurations involving both S and AS
states, loses its stability as the inter-dot barrier is increased.

\begin{figure}
\centering
\includegraphics[clip,angle=0,width=0.45\textwidth]{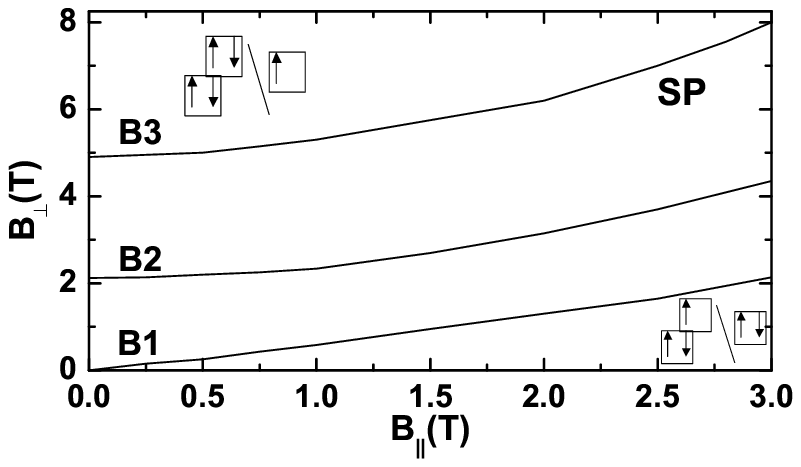}
\caption{$N=5$ stability diagram in the
$(B_{\parallel},B_{\perp})$ plane for \bone, \btwo, and \bthree,
as obtained within the SP non-interacting
scheme.} \label{fig12}
\end{figure}

\begin{figure}
\centering
\includegraphics[clip,angle=0,width=0.45\textwidth]{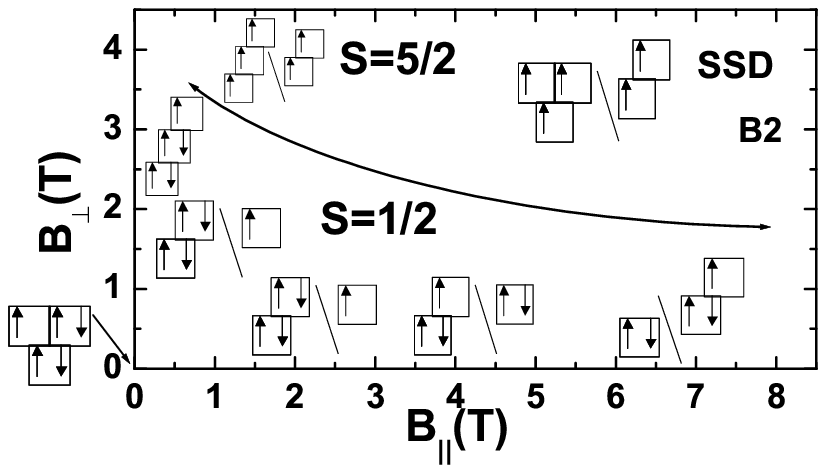}
\caption{$N=5$ stability diagram in the
$(B_{\parallel},B_{\perp})$ plane for \btwo, calculated within the SSD
scheme.} \label{fig13}
\end{figure}

\begin{figure}
\centering
\includegraphics[clip,angle=0,width=0.5\textwidth]{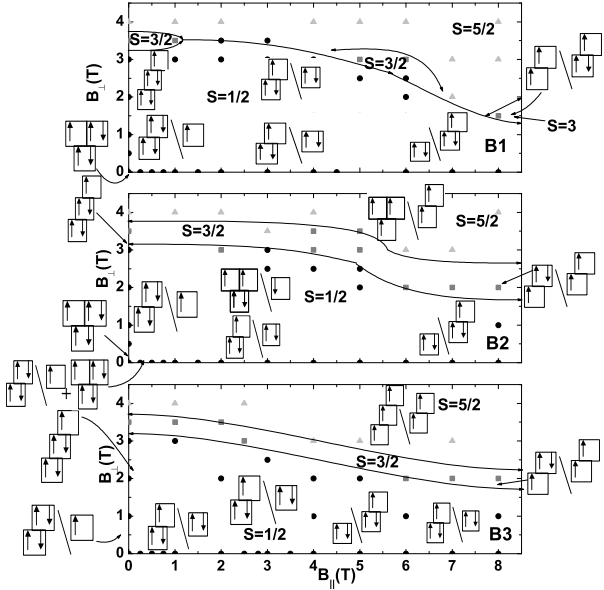}
\caption{$N=5$ stability diagram in the
$(B_{\parallel},B_{\perp})$ plane for \bone, \btwo, and \bthree, as predicted by
the FCI calculations. Gray triangles: $S=5/2$ GS; gray squares: $S=3/2$ GS;
black circles: $S=1/2$ GS. Solid lines are guides to the eye. Insets:
main components of the GS wave functions (see text).} \label{fig14}
\end{figure}

Similar considerations apply to the $N=5$ case (see
Figs.~\ref{fig12} ,\ref{fig13}, and \ref{fig14} for SP, SSD, and FCI
schemes, respectively). Typically, several transitions occur in the
GS within the same spin configuration. For example, along the
$B_{\parallel}$ axis of Fig.~\ref{fig14} several configurations
occur, reflecting the crossing between $p$ levels in the S and AS
mini-bands (see Fig.~\ref{fig2}). As the barrier $L_{b}$ is
decreased, the splitting between $s$ and $p$ orbitals becomes of the
same order of magnitude of $\Delta_{\mathrm{SAS}}$, and the number
of transitions in the GS increases, too. Also in this case a SSD
approach is inappropriate. As in the $N=4$ case, we have an
intermediate phase ($S=3/2$) between the unpolarized ($S=1/2$) and
the completely polarized ($S=5/2$) ones, which is only stable in the
intermediate coupling regime; this phase is absent in the SSD
results (Fig.~\ref{fig13}).

The above results were obtained by keeping $\hbar\omega_0$ fixed
with respect to $N$. In single-electron tunneling experiments,
however, $\hbar\omega_0$ changes with the gate voltage, in order to
inject (extract) electrons into (from) the AM.\cite{rontani} A
relation linking $\omega_0$ with $N$ may be derived by imposing that
the electron density is approximately constant during the charging
processes, $\omega_0 \propto N^{-1/4}$
(Refs.~\onlinecite{reimann1,rontani}). This allows us to
qualitatively understand what changes the AM phases undergo if we
try to mimic the dependence of $\omega_0$ on $N$. Indeed, if
$\hbar\omega_0$ decreases as $N$ increases, the ratioes of this
energy to the other relevant energy scales ---namely
$\Delta_{\text{SAS}}$, $\hbar\omega_c$, and Coulomb matrix elements,
respectively--- decrease as well. This implies that: (i) the AM
phase boundaries in the $(B_{\parallel},B_{\perp})$ plane move in
order to ``squeeze'' the stability regions of ``ionic'' phases (ii)
the transition to the MDD phase along the $B_{\perp}$ axis is faster
(iii) correlation effects are stronger.

Finally, we note that the present results for $N=4,5$ at
$B_\parallel=0$ are in agreement with the ones shown by Rontani
\emph{et al.}~dealing with similar samples,\cite{rontani} except for
little discrepancies due to different parameters used in the
calculations, in particular the band-offset $V_0$ and the lateral
confinement energy $\hbar\omega_0$.

\section{Conclusions}

We have investigated the relative stability of quantum phases of
interacting electrons in AMs under a magnetic field. Such phases can
be identified, e.g., in transport experiments in the Coulomb
blockade regime, where the addition energy is
measured.\cite{tarucha1} A very sensitive test of few-electron
phases is non-linear transport through charged QDs, where a few
excited states in a small energy window are
accessed.\cite{kou97,Ota05,rontani} In addition to predict the
stability regions of the different spin phases, this work confirms
that correlation effects play a fundamental role in determining the
physics of the system, and a simple picture neglecting these effects
could fail in interpreting the relevant experiments.

In particular, we have
compared FCI predictions with (i) a SP picture, neglecting all
interactions, (ii) a SSD scheme, where only direct and exchange
interactions between charge carriers are taken into account. These
comparisons allowed us to identify the effects of correlations,
particularly those induced in the coupled structure by an in-plane
magnetic field, in analogy to what happens in single QDs with an
applied vertical field. In many respects, only the exact
diagonalization of the few-body Hamiltonian is able to provide a
correct description of correlated states.

We thank Seigo Tarucha, Shinichi Amaha, Andrea Bertoni, and Filippo
Troiani for helpful discussions. This work has been supported in
part by the Italian Ministry of Foreign Affairs (Ministero degli
Affari Esteri, Direzione Generale per la Promozione e la
Cooperazione Culturale), Italian Ministry of Research
(FIRB-RBIN04EY74), and the CINECA-INFM
Supercomputing Projects 2005 and 2006.


\begin{thebibliography}{99}
\bibitem{nota-labelling}
Here we use the group-theoretical labelling at $B_\perp =0$ for convenience;
$A_g$, $B_u$, \ldots indicate SP levels of increasing energy, calculated
acording to Sec.~\ref{sub:Single--states} assuming no symmetry.
\bibitem{nota-s1}
Only for strongly coupled AMs at $B=0$, when the tunneling gap
$\Delta_{\mathrm{SAS}}$ exceeds the $s-p$ separation $\hbar\omega$,
the triplet configuration, obtained filling the two S $p$ orbitals
with one electron each, is stable.
\end{thebibliography}
\end{document}